\documentclass[graphics,showpacs,noshowkeys]{elsart}
\begin{document}
\input{epsf.sty}
\begin{frontmatter}
\title{Weak turbulence and collapses in the 
Majda-McLaughlin-Tabak equation: Fluxes in wavenumber 
and in amplitude space}
\author[Chemnitz]{Benno Rumpf \thanksref{corresp}}, 
\author[Dresden]{Laura Biven \thanksref{Newyork}}
\address[Chemnitz]{Institute of Physics,
Technical University of Chemnitz\\
09107 Chemnitz, Germany}
\address[Dresden]{Max-Planck-Institute for the Physics of Complex Systems, \\
N\"othnitzer Stra\ss e 38, 01187 Dresden, Germany
}
\thanks[corresp]{Corresponding author.\\
E-mail address: benno.rumpf@physik.tu-chemnitz.de}
\thanks[Newyork]{Permanent address: Bard HS Early College, 
525 East Houston St., New York, 
NY, 10002, USA}
\begin{abstract}
The turbulent energy flow of the one-dimensional Majda-McLaughlin-Tabak 
equation is studied numerically. The system exhibits weak turbulence 
for weak driving forces, while weak turbulence coexists with  
strongly nonlinear intermittent collapses when the system is strongly 
driven. These two types of dynamics can be 
distinguished by their energy and particle fluxes. 
The weakly turbulent process can be characterized by 
fluxes in wavenumber space, while additional fluxes in amplitude 
space emerge in the intermittent process. The particle flux is directed 
from low amplitudes towards high amplitudes, and the energy flows in 
the opposite direction. 
\end{abstract}
\end{frontmatter}
\section{Introduction}
This paper investigates two different types of turbulence, 
firstly, the energy transport of weakly interacting dispersive 
waves with low amplitudes \cite{zlf}, secondly, an intermittent 
type of turbulence \cite{bnn} 
where localized high-amplitude structures appear 
from a low-amplitude background, and decay rapidly \cite{ksz}
\cite{dnpz}\cite{rn2}. 
Important cases of weak turbulence 
have been found in the dynamics of Langmuir waves in 
plasmas \cite{mrz}, capillary \cite{hal} and gravity waves \cite{dkz} 
in hydrodynamics, in magnetohydrodynamics \cite{gal}, 
and in Bose-Einstein condensates \cite{lnwj}. 
Collapses in Langmuir-waves \cite{z72} and self-focusing effects in optics 
\cite{dnpz},\cite{kgz} 
are examples for the intermittent kind of behavior. \\
Turbulent energy flows emerge when such dynamical systems are driven 
on a long spatial scale, and damped on a short viscous scale \cite{zlf}. 
Models of turbulent flows assume that the dynamics is almost Hamiltonian 
in the inertial range between the damping range 
and the driving range, 
and energy is transported from sources to sinks through this 
window of transparency with no damping or driving. 
Symmetries besides the time-invariance 
are related to conserved quantities in addition to the 
Hamiltonian. Examples for such quantities are 
the light intensity in nonlinear optics \cite{kgz} or 
the average momenta in hydrodynamics \cite{saf}. 
When supplied and dissipated by driving and damping, 
there are fluxes also of these quantities. 
Fluxes that are directed from a source both toward long and 
toward short length scales 
can be modeled with an additional sink at small wavenumbers. 
\\ 
The fluxes of the conserved quantities determine 
the statistically stationary nonequilibrium of the Hamiltonian 
dynamics in the inertial range. 
The theory of weak wave turbulence computes the nonequilibrium state 
of weakly interacting dispersive waves, depending on boundary conditions 
for the fluxes. 
A closed kinetic equation for the 
two-point functions can be derived if, with the existence of a small
parameter, the nonlinear coupling is much smaller than the linear
dispersive effects, so that higher order correlations can be reduced
to powers of the two-point function. The kinetic equations 
can be solved analytically. This yields the Kolmogorov-Zakharov 
spectrum and formulas for the fluxes \cite{zlf}. 
The theory predicts that the system will have close to Gaussian
statistics even after the weak nonlinearity is relevant.
The intermittent occurrence of high-amplitude structures \cite{rn} violates 
the assumptions of weak turbulence, as these 
structures indicate strongly coherent waves. Consequently, the closure 
of the kinetic equation is not possible, and there is a strongly 
nonlinear interaction. 
\\
In recent years massive numerical
experiments have been carried out 
\cite{mmt}, \cite{cmmt}, \cite{zgpd}, \cite{zdp} to 
check the assumptions of the 
Gaussian closure of the kinetic equations. This was partly thanks 
to the development of dynamical system by Majda, McLaughlin, and 
Tabak ('MMT' equation) \cite{mmt},\cite{cmmt}, 
which we use in its most simple focusing version 
\begin{equation}
i\frac{\partial \phi}{\partial t}=
|\frac{\partial}{\partial x}|^{1/2}\phi- \phi|\phi|^2
+\mathcal{F}+\mathcal{D}.
\label{mmteq}
\end{equation} 
$\phi$ is a complex field that depends on one spatial variable $x$. 
The fractional derivative $|\partial/\partial_x|^{1/2}\phi$ stands for the 
Fourier-transform of the term $\sqrt{|k|} a_k$ with the mode 
$a_k=(1/L)\int \phi(x)\exp{(ikx)} dx$, or equivalently, for the 
inverse of a fractional integral \cite{so}. 
The concave dispersion $\omega(k)=\sqrt{|k|}$ 
yields more four wave resonances than a convex dispersion law 
(such as the dispersion $\omega(k)\sim k^2$ of the nonlinear 
Schr\"odinger equation) does. This leads to 
an efficient energy transport even in only one dimension, and allows 
numerical simulations in a wider range in the wavenumber space compared 
to simulations in two or three spatial dimensions. 
The general MMT equation has fractional derivatives of any order  
both in the linear and in the 
nonlinear part, but the semiderivative 
and the simple nonlinearity of (\ref{mmteq}) is by the signs of 
the fluxes \cite{zgpd} suitable  
for the study of wave turbulence. 
\\
$\mathcal{D}$ is a damping term that affects very short waves and very 
long waves only. The driving force $\mathcal{F}$ affects waves on an 
intermediate 
length scale. The length scale of the driving force and the length scale 
of the short-wave damping are widely separated by the window of 
transparency where no damping or driving occurs. \\
The MMT equation for zero driving $\mathcal{F}$ and damping 
$\mathcal{D}$ derives from the Hamiltonian (or "energy") 
$E=E_2+E_4$, with the quadratic coupling part 
$E_2=\int ||\partial/\partial x|^{1/4}\phi|^2 dx =\sum \sqrt{k} |a_k|^2$, 
and the nonlinear part $E_4=-\int |\phi|^4/2 dx$. 
The wave action or "particle number" 
$n=\int |\phi|^2 dx=\sum |a_k|^2$ is a second conserved quantity 
which is related to the system's phase symmetry. \\
Numerical simulations of (\ref{mmteq}) found the Kolmogorov-Zakharov
spectrum as predicted by weak turbulence theory, 
but also concluded that most 
of the energy transport is related to strongly nonlinear collapses 
\cite{zgpd}, which 
coexist with weakly interacting low-amplitude waves. In this case, 
high-amplitude peaks emerge in a very similar way as in the 
corresponding Hamiltonian systems where this process maximizes 
the entropy \cite{lr}. 
Pure wave turbulence for the MMT equation has been found for freely decaying 
turbulence \cite{cmmt}, but not in the stationary nonequilibrium. 
A steeper spectrum than predicted by wave turbulence has been obtained 
for the defocusing MMT equation (where the sign of the nonlinearity 
is opposite to the focusing equation (\ref{mmteq})) 
and in some cases for the focusing equation \cite{cmmt}. 
This lead to the proposal of an alternative closure from which 
the steeper spectrum can be derived \cite{mmt}. 
\\
The aim of this paper is to study both the weakly turbulent 
state and the collapse dynamics by measuring the fluxes of 
energy and particles. In fact, as we show here, flux measurements are a
more reliable indicator of the transport dynamics than the energy spectrum.
We compute energy and particle fluxes by measuring 
the input and output of these quantities that is caused 
by the driving force and the damping in our numerical experiments. \\
We give evidence that a pure weakly turbulent process emerges if 
the driving forces are sufficiently weak. 
The fluxes in wavenumber space correspond to the predictions 
of weak turbulence theory (section 3). 
\\
For stronger external driving we find  
a transition to a process where weak turbulence and 
collapse dynamics coexist. 
Despite the fact that the formation 
of high-amplitude structures is not in agreement with the assumptions 
of weak turbulence our simulations of (\ref{mmteq}) find the weak
turbulence spectrum. 
The collapse dynamics is characterized 
by fluxes of energy and particles both in wavenumber and in amplitude space. 
These fluxes in amplitude space contain information on the 
coherent process (section 4). 
\\
We present numerical results for one simulation that is representative 
for weak driving and damping (Figs. 2,3,4) and one for strong driving 
and damping (Figs. 2,6,7,8). Figs. 9,10 show data from another simulation 
with a medium power driving force. These two figures show time-dependent 
quantities, all other figures show time-averaged quantities. 
Figs. 5,11 show results of a series 
of simulations with various driving forces. 
\section{Numerical method}
We integrate equation (\ref{mmteq}) numerically using a pseudospectral method 
with an integrating factor similar to the method described in \cite{mmt}. 
$\phi(x)$ is defined on the interval $[0,4096]$ with periodic 
boundary conditions. For its integration we use $L=4096$ modes, 
with wavenumbers $k$ in the Brillouin zone $]-\pi,\pi]$. 
The linear part can be eliminated analytically using an integrating factor,
and the resulting non-stiff equation is integrated with 
a multistep (Adams) method. \\
We use driving and damping forces that act as time-periodic 
$\delta$-kicks, while the system is governed by the Hamiltonian 
MMT equation for one time unit between the subsequent kicks. 
The external driving force $\mathcal{F}$ is applied to the 
modes in the interval 
$20<|k|L/(2\pi)\le 40$. 
The driving $\delta$-kicks increase 
the amplitude as $a_k\rightarrow (1+\lambda)a_k$ 
with $\lambda=0.1$. This driving force is only applied to modes with an 
amplitude below a threshold $|a_k|^2<c_{0}/k$, so that the total energy 
input reduces when more modes exceed this threshold. \\
The damping $\mathcal{D}$ is applied both at low wavenumbers 
($|k|L/(2\pi) \le 20$) and at high wavenumbers $\pi/2\le|k|$, while  
modes in the window of transparency 
$40 < |k|L/(2\pi) < L/4$ are neither damped and nor driven. 
The broad high-$k$ damping minimizes  
the aliasing error of the numerical computation. 
Ideally, the low-$k$ damping and driving range 
should also be chosen larger, but this would narrow the window 
of transparency. 
The damping $\delta$-kicks  
decrease the amplitude as $a_k\rightarrow (1-\lambda)a_k$ 
with $\lambda=0.1$. This damping is applied only to modes in the 
high-$k$ window when the amplitude exceeds the threshold $|a_k|^2>c_+/|k|$, 
and to low-$k$ modes above the level $|a_k|^2>c_-$. 
This type of damping becomes more effective when more modes exceed this 
threshold. 
\\
The purpose of this type of damping and driving is purely practical: 
Damping and driving by time-periodic $\delta$-kicks has the 
advantage that the input and output of energy and particles 
can be measured 
\begin{figure}
\epsfbox{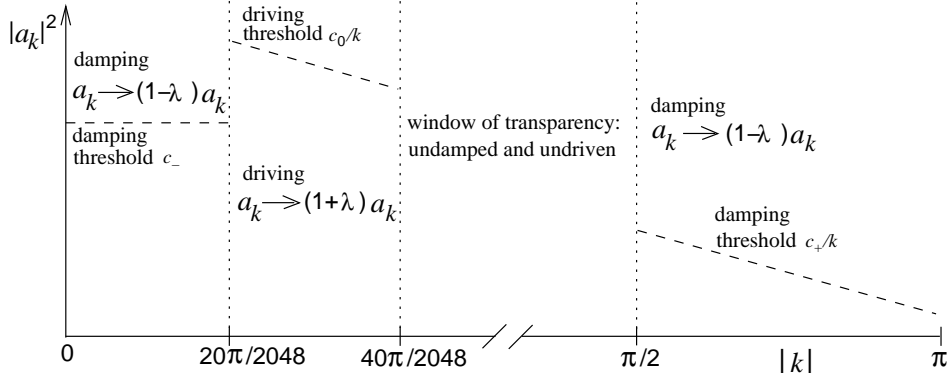}
\caption{The driving force acts on modes $a_k$ 
with $20 < |k|L/2\pi \le 40$ when $|a_k|^2$ is below a threshold 
$c_{0}/k$. 
The driving $\delta$-kicks increase these modes 
as $a_k\rightarrow (1+\lambda)a_k$ 
with $\lambda=0.1$. 
The damping decreases the modes of the interval $|k|L/2\pi \le 20$ 
as $a_k\rightarrow (1-\lambda)a_k$ 
when $|a_k|^2$ above a constant 
threshold $c_-$, as well as the modes in the interval 
$\pi/2 < |k| \le \pi$ when 
$|a_k|^2$ is above a threshold $c_+/k$.}
\end{figure}
directly, and are distinguishable from the Hamiltonian dynamics of
(\ref{mmteq}). The effect of the kicks  
is quite small in all our simulations, so that there is little difference 
between snapshots before and after the kicks. 
The damping is restricted to the two windows in $k$-space 
in order to keep the window of transparency 
free of dissipation. 
The threshold parameters $c_-$, $c_{0}$ and $c_+$ allow us to control 
the amplitudes in the driving and the damping regions. In particular, 
the driving force can be increased by increasing the threshold $c_0$.  
This type of forcing is similar to coupling the system 
to heat baths with different temperatures in the driving and damping 
regions. It appears that this leads to a relatively quick relaxation 
to a statistically stationary state. In this state, it is equivalent 
to other types of forcing that cause the same fluxes. In our simulations 
we choose $c_-=c_0/60$ and 
$c_{0}$ slightly larger than $ c_+$. This choice has turned out to yield 
Kolmogorov-Zakharov spectra with relatively small deviations at the boundaries 
of the window of transparency. \\
For comparison, we have performed 
simulations with time-continuous driving and hyperviscous 
damping, which show no significant differences to the $\delta$-driven 
case. Our results are 
also in agreement with those reported in the literature. 
In particular, we have found WT spectra \cite{zgpd} for the 
focusing MMT equation (\ref{mmteq}), while we find the steeper 
MMT spectrum \cite{mmt} for its defocusing version. 
The results in this paper are robust under various modifications 
of the scheme. We have performed simulations with various 
resolutions from 1024 modes to 8192 modes, and various types of damping and 
driving. Other modifications include the ranges in $k$-space where 
damping and driving is 
applied, the time intervals between the kicks and the strength of the kicks. 
We have tested of the integration scheme for undamped and undriven 
equations by simulating  
the well-known solitons of the nonlinear 
Schr\"odinger equation, and by checking the conservation of 
energy and particles by the MMT equation. 
\\
In all simulations the system is first allowed to relax until 
it reaches an almost stationary nonequilibrium state, which 
takes $10^4-10^5$ time units depending on the particle input. 
Time-averaged quantities such as power spectra and energy fluxes are computed 
for the nonequilibrium state in a second round of integration. 
In this state, 
the input of particles and energy matches the output, and the 
energies $E_2$ and $E_4$ and particle number $n$ 
are constant with only small fluctuations. 
The integration time is long enough 
to average away the effect of these random fluctuations. 
\section{The weakly turbulent state}
\subsection{Power spectrum and amplitude statistics}
The dynamics of (\ref{mmteq})  depends critically on the strength 
of the driving force and the damping, and we study the dynamics 
in two representative 
\begin{figure}[b]
\epsfbox{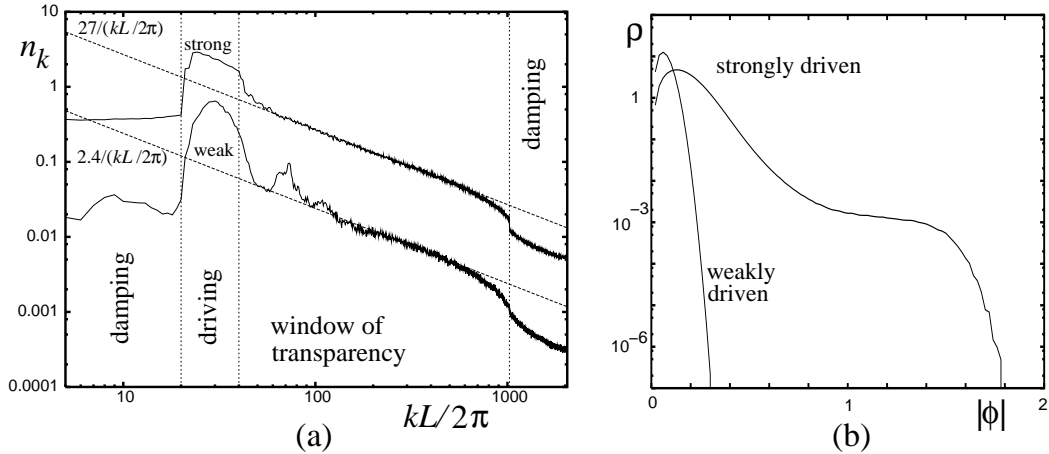}
\caption{(a): The time-averaged wave-action 
$n_k=<|a_k|^2+|a_{-k}|^2>$ ($n=\sum n_k$ is the total particle number) of the 
MMT equation  exhibits Kolmogorov-Zakharov spectra  
$n_k\sim k^{-1}$ in the window of transparency both for weak and 
for strong driving forces. 
The dotted lines are fits 
$n_k=c/(kL/2\pi)$ with $c=2.4$ and $c=27$ of 
the spectra in the window of transparency. 
(b): Probability density of the amplitude $|\phi|$ for the 
weakly and the strongly driven system. The probability for high amplitudes 
decays as $\sim \exp (-|\phi|^2)$ for the weakly driven system. 
The shoulder-shaped distribution for the strongly driven 
system exhibits a significant probability for  
high-amplitude structures ($|\phi|\approx 1.5$).
Particle number and energy fluctuate about constant values in both 
simulations 
(weak driving: $n\approx 18$, $E_2\approx 7.5$, $E_4\approx -0.08$, 
strong driving: $n\approx 150$, $E_2\approx 75$, $E_4\approx -10$). 
}
\end{figure}
simulations, one for weak driving, and one for 
strong driving. Fig.2(a) shows the 
time-averaged wave action $n_k=<|a_k|^2+|a_{-k}|^2>$ over the wavenumber. 
Both for the weakly and the strongly driven system, we obtain 
a Kolmogorov-Zakharov spectrum $n_k=c/(kL/2\pi)$ in the 
window of transparency. The coefficient $c$ 
measures the particle number for waves in the window of transparency, 
which is $c=2.4$ for the weakly driven system and 
$c=27$ for the strongly driven system. 
In the regions in $k$-space 
where the driving (damping) is applied, $n_k$ is higher (lower) 
than the level of the spectrum $n_k=c/(kL/2\pi)$. 
In the weakly driven system, 
the spectrum has also two humps at harmonics of the driving force at 
the low-$k$ end of the window of transparency. 
These effects could be reduced with a driving force that acts on 
a broader $k$-band. 
\\
In Fig.2(b), $\rho(|\phi|) d|\phi|$ gives the probability that 
the amplitude at one location is between $|\phi|$ and $|\phi|+d|\phi|$. 
The distribution of the amplitudes is qualitatively different 
for the two simulations. 
In the weakly driven system we find a distribution 
$\rho\sim |\phi|exp(-|\phi|^2)$ without any incident of an amplitude 
$|\phi|>0.3$ during $5\cdot 10^5$ integration cycles. The probability density 
of the strongly driven system is again gathered at small $|\phi|$, 
but its shoulder-like shape shows that amplitudes of the order of one 
are obtained with a probability of about $10^{-3}$.  
\subsection{Input and output of particles and energy}
Fig.3 shows the input and output rates of particles and energy by the driving 
and damping forces as functions of $k$ for the simulation with weak 
driving of Fig.2. 
For clarity, low wavenumbers (a) 
and high wavenumbers (b) are plotted on different scales. 
\begin{figure}[b]
\epsfbox{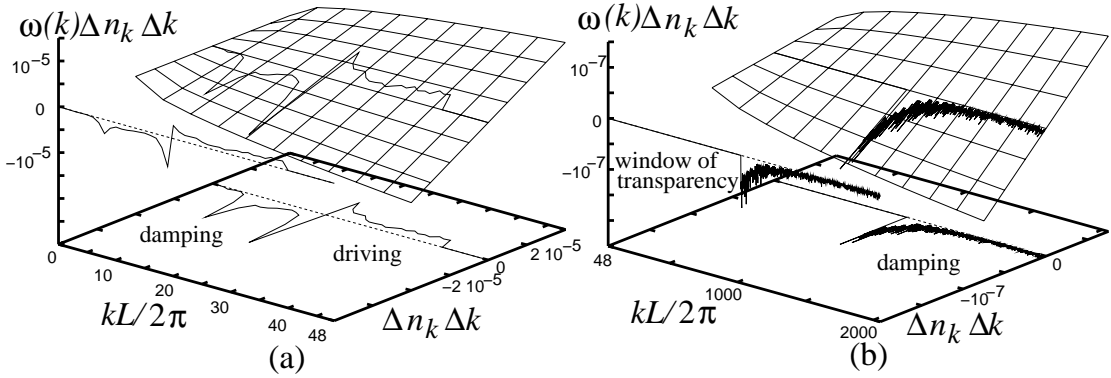}
\caption{Particle- and energy-balance for the weakly driven 
system of Fig.2. 
The driving and damping forces change the particle contents 
of the mode $k$ by $\Delta n_k\Delta k$ per time unit. 
$\Delta n_k$ is the change of the particle density, $\Delta k=2\pi/4096$ 
is the distance between adjacent modes. 
This particle input or 
output is connected to a change of the coupling energy by 
$\omega(k)\Delta n_k\Delta k$. (a): The low-$k$ damping range and 
driving range. 
(b): The window of transparency and the high-$k$ damping range. 
}
\end{figure}
Positive $\Delta n_k$ show the input of particles in the driving range (Fig.3(a)), 
while negative $\Delta n_k$ indicate particle losses both in the low-$k$ 
(Fig.3(a)) and in the high-$k$ damping ranges (Fig.3(b)). 
The total particle number is unchanged since 
input and output of particles match as $\sum\Delta n_k\Delta k=0$, 
and the particle number is also conserved by the Hamiltonian dynamics. 
\\
A change $\Delta n_k$ of the particle density at the mode 
$k$ also changes the density of the coupling energy 
$E_2$ at this  
wavenumber by $\omega(k) \Delta n_k\Delta k$. Consequently, the gains and 
losses of particles by the driving and 
damping translate to gains and losses 
of energy via the coefficient $\omega(k)=\sqrt{k}$. 
The energy gain in the 
driving range matches the energy loss in the two damping ranges, 
so that $\sum \omega(k)\Delta n_k\Delta k=0$. The influence 
of driving and damping on the nonlinear 
energy $E_4$ is very small.  
\subsection{Particle flux of the weakly driven system}
In a statistically stationary nonequilibrium state, the particle number 
$|a_k|^2+|a_{-k}|^2$ at $|k|$ fluctuates about a 
time-independent average value 
$n_k$ (Fig.2a). 
Particles that are injected in the driving range flow 
to the two damping regions, where particles are removed at the same rate. 
This defines two fluxes of particles, one in a direct cascade through 
the window of transparency (where no particles are injected or removed) 
to the high-$k$ damping, and one in an inverse cascade to the low-$k$ 
damping range. 
\begin{figure}[b]
\epsfbox{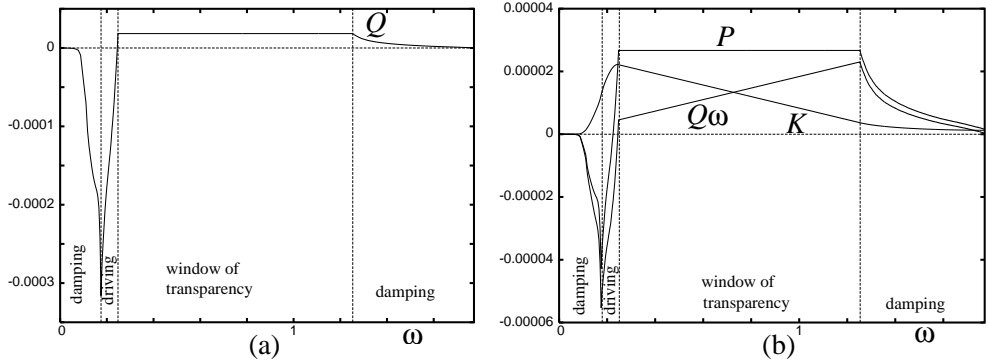}
\caption{Particle- and energy-fluxes in $\omega$-space 
for the weakly driven system 
of Fig.2. (a): The particle flux $Q$ is positive (i.e. particles 
move toward high $\omega$) in the window of transparency and in the 
high-$\omega$ (or high-$k$) damping range, and it is strongly negative 
in the low-$\omega$ damping range and in most 
of the driving range. Its gradient is negative in the damping 
ranges, where particles are removed from the system, and positive 
in the driving range, where particles are injected into the system. 
$Q$ is constant in the window of transparency. It is zero at 
$\omega=0$ and at $\omega=\sqrt{\pi}$, as the particle input matches the 
particle output. 
(b): The energy flux $P$ has the same qualitative characteristics as 
$Q$. $Q\omega$ is the energy drift 
of the particle flux. $K=P-Q\omega$ is the energy transfer 
from the inverse cascade at low $\omega$ (where energy is released) to 
the direct cascade where energy is absorbed by the energy drift.  
}
\end{figure}
The particle flux $q(k)$ through $k$-space 
follows from the continuity equation for 
the particles 
\begin{equation}
\Delta q(k) / \Delta k - \Delta n_k / \Delta t =0
\label{qflux}
\end{equation}
where $\Delta k=2\pi/4096$ and $\Delta t=1$ in our simulations 
(a corresponding partial differential equation $\Delta\rightarrow \partial$ 
applies to continously damped and driven, infinite systems). 
$\Delta n_k$ is the particle input or output by the 
external driving and damping, which is minus the 
particle change that is caused by the flux $q(k)$. 
This definition yields the minus sign in equation (\ref{qflux}). 
The particle flux is obtained as 
$q(k)=\sum_{\tilde{k}=0}^{k}\Delta n_{\tilde{k}}\Delta\tilde{k}/\Delta t$. 
As the system's 
particle number is constant, 
the particle flux vanishes at $k=0$ and at the short-wave boundary 
of the Brillouin-zone $k=\pi$. 
The flux can be defined in frequency 
space as $Q(\omega)=q(k(\omega))$ which has similar properties 
since $\omega(k)=\sqrt{|k|}$ is bijective in the interval $[0,\pi]$. 
Fig.4 shows the particle flux $Q(\omega)$ for the simulation of 
Fig.3. A negative sign 
indicates a flux toward small frequencies (or wavenumbers), a positive sign 
shows a flux to high $\omega$. 
The flux is strongly negative in the the inverse cascade 
at low $\omega$. The sign changes at some point 
in the driving region, and in the window of transparency 
there is a small positive flux to high $\omega$. 
The slope is positive 
in the driving range, and negative in the damping ranges. 
The flux of this direct cascade is $k$-independent in the window of
transparency, and decays in the damping region at $k>\pi/2$ where 
particles are absorbed. 
\subsection{Energy flux in the weakly driven system}
Similarly to the particles, the energy that is injected at the driving 
scale flows to the two damping ranges. 
The energy flux 
$p(k)=\sum_{\tilde{k}=0}^{k}
\omega(\tilde{k})\Delta n_{\tilde{k}}\Delta\tilde{k}/\Delta t$ 
can be defined from the energy continuity equation 
\begin{equation}
\Delta p(k) / \Delta k - \omega(k)\Delta n_k / \Delta t =0
\label{efluxeq}
\end{equation}
where the energy sources and sinks are located at the same wavenumbers 
as the particles sources. 
The constant energy of the nonequilibrium state requires 
that there is no flux through the boundaries 
$p(0)=p(\pi)=0$. 
Fig.4(b) shows the corresponding flux in $\omega$-space 
$P(\omega)=p(k(\omega))$. \\
As particles move toward higher values of $\omega$ in the direct cascade, 
their load with coupling energy increases proportionally to $\omega$. 
The energy transport of the particle flux is measured by the quantity 
$Q(\omega)\omega$, which we call the energy drift. 
The energy drift depends on $\omega$ even in regions that are 
remote from energy sources and sinks:  
In the window of transparency, it increases with $\omega$  
while the energy flux $P(\omega)$ and the particle flux $Q(\omega)$ are
constant. The inverse cascade is linked to a negative energy drift that 
approaches zero for $\omega\rightarrow 0$. 
\\
$K=P-Q\omega$ amounts for the discrepancy of the energy flux and the 
energy drift. It measures the part of the energy flux $P$ that is 
not related to the particle flux. We call $K$ the energy transfer term, 
because it describes the transfer of energy from the inverse cascade  
to the direct cascade: Particles of the inverse cascade release energy 
as they move towards lower $\omega$, which feeds the energy transfer process. 
The energy transfer $K$ shifts this energy toward 
higher frequencies by four-wave interactions. This energy is absorbed 
by the energy drift in the 
direct cascade, where the particles demand energy in order to move to higher 
$\omega$. 
In other words, the inverse cascade drives the direct cascade 
via the energy transfer. $K$ is positive for all wavenumbers in this 
simulation, and vanishes at the boundaries. \\
Fig.4 and Fig.2 give evidence for a pure weakly 
turbulent flow without strongly nonlinear processes. 
The amplitudes are always small, 
and the coupling part of the Hamiltonian carries almost all energy.  
The nonlinearity acts only as a 
weak coupling force of linear waves, without directly interfering in the 
energy input or output. 
As suggested in \cite{zgpd}, 
the main ingredient to obtain such a solution numerically is a  
sufficient number of modes in the low-$k$ damping range and in the 
driving range. This avoids a bottleneck in the particle flow of the 
inverse cascade due to the lack of resonant modes. \\
The Zakharov solution $n_k\sim 1/k$ to the kinetic equation involves 
a finite energy 
flux $P>0$ and a zero particle flux $Q=0$ in the direct cascade. 
This may be regarded as the limit of the fluxes of Fig.4 for an infinite 
extension of the window of transparency, with a high-frequency damping 
at $\omega\rightarrow\infty$. 
The energy transfer $K$ is again fed by the release of energy 
in the inverse cascade, but it is qualitatively different in the direct 
cascade. 
The Zakharov solution has a zero energy drift in the direct cascade, so 
that the energy transfer equals the energy flux $K=P$, and 
is $\omega$-independent in the window of transparency. 
\subsection{Transition of the particle flux}
We now determine how the strength of the particle flux changes with 
the particle number. We run a number of simulations 
with various driving powers and determine  
the coefficient $c$ of the spectrum $n_k=c/(kL/2\pi)$ 
for each of these simulations. $c=2.4$ of Figs.3,4 
corresponds to the weakest driving force. 
Fig.5 shows the strength of the particle fluxes $Q$ 
of the direct cascade and the 
inverse cascade for various driving forces. 
$Q_+$ is the particle flux in the window of transparency, where the 
particle flux of the direct cascade reaches its maximum. 
$|Q_-|$ is the modulus of the particle flux at the boundary between the 
low-$k$ damping range and the driving range (see Figs.3(a),4(a)). 
Both fluxes increase with $c$, 
and the lines in Fig.5(a) are tentative interpolations $Q\sim c^{3.45}$. 
\begin{figure}
\epsfbox{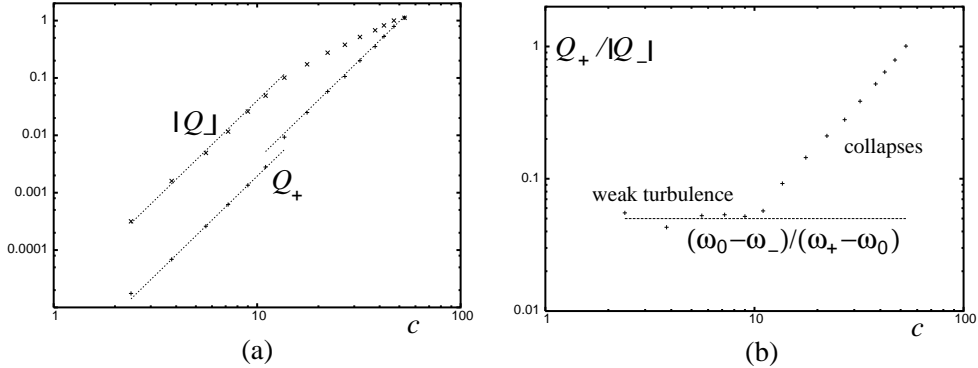}
\caption{(a): Maximum particle flux of the direct cascade $Q_+$ and 
of the inverse cascade $|Q_-|$ versus the coefficient $c$ of the 
wave action $n_k=c/(kL/2\pi)$. The lines give a fit with 
$\sim c^{3.45}$. (b):  
Ratio of direct and inverse cascade versus the wave action. 
For low particle densities, the ratio agrees to the prediction 
of weak turbulence theory 
$Q_+/|Q_-|=(\omega_{0}-\omega_{-})/(\omega_{+}-\omega_{0})\approx0.05$, 
where $\omega_+\approx 1.5$ is the average frequency of high-$k$ damping 
and $\omega_-\approx 0.15$ of low-$k$ damping, while 
$\omega_{0}\approx 0.21$ is the 
frequency where the driving occurs. 
}
\end{figure}
For comparison, the wave turbulence closure gives 
$Q\sim c^{3}$ for four-wave interactions.\\ 
The input of particles  and of energy 
matches the output only for a certain 
ratio \cite{zlf} of the strengths of the fluxes $Q_+$ and $Q_-$. 
We assume for simplicity that the driving force acts only at one 
effective frequency $\omega_0$, and that the 
high- and low-frequency damping act 
at $\omega_+$ and $\omega_{-}$ respectively. 
The particle output rates at these sinks 
are $Q_+$ and $|Q_-|$, with corresponding energy output rates 
$Q_+\omega_+$ and $|Q_-|\omega_-$. Input of particles $Q_++|Q_-|$ 
and of energy $(Q_++|Q_-|)\omega_0$ and their output match if 
$Q_+/|Q_-|=(\omega_{0}-\omega_{-})/(\omega_{+}-\omega_{0})\approx0.05$, 
where we have set $\omega_+\approx 1.5$, $\omega_-\approx 0.15$, 
$\omega_{0}\approx 0.21$. 
Fig.5(b) shows this ratio of the two fluxes. For weak driving forces 
(corresponding to $c<10$), $Q_+$ and $|Q_{-}|$ both increase at the same rate
with $c$, so that we find the expected fixed ratio of fluxes 
$Q_+/|Q_-|\approx0.05$ (line in Fig.5(b)). 
Above $c\approx 10$ however, the flux $|Q_-|$ increases at a lower rate 
with $c$, and the ratio $Q_+/|Q_-|$ increases. This is obviously not
compatible with the assumption of a weakly turbulent state. 
\section{Collapses and fluxes in amplitude space}
\subsection{Particle flux and energy flux for the strongly driven system}
Above the transition point of Fig.5(b) 
we observe intermittent high-amplitude structures. 
These structures are visible in the amplitude statistics 
of Fig.2(b), but not in the power spectrum Fig.2(a) since only few 
of the particles contribute to collapses, and their Phillips 
spectrum decays more rapidly with $k$ than the WT spectrum 
of the low-amplitude background \cite{zdp}. 
We focus on the fluxes of particles and energy related to the strongly 
nonlinear events, and mention their dynamical properties only briefly. \\
Figs.6-8 show results for the strongly driven system of Fig.2 with $c=27$. 
Fig.6 shows the fluxes of particles and energy in wavenumber space. 
\begin{figure}[t]
\epsfbox{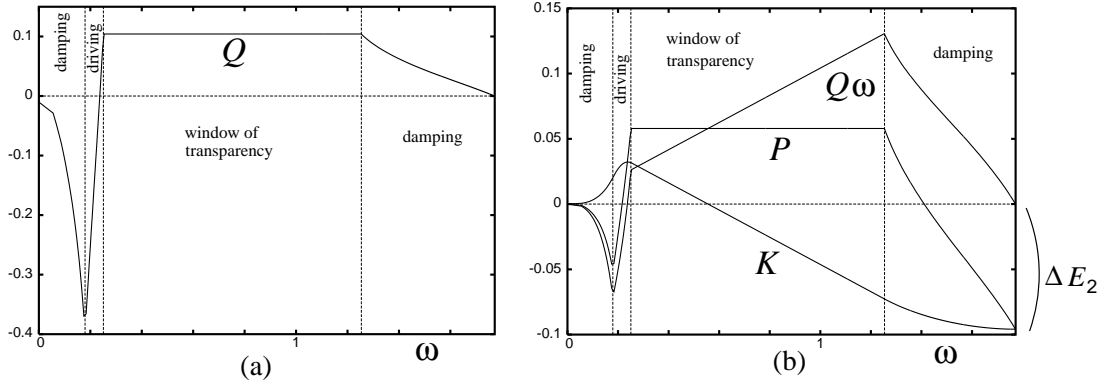}
\caption{Particle- and energy-fluxes of the strongly driven system 
of Fig.2. (a): The particle flux $Q$ has the same shape as for the 
weakly driven system (Fig.3), but the positive flux towards high $\omega$ 
has a higher relative strength. Again there is no net gain or loss 
of particles. (b): Unlike for the weakly driven system (Fig.3b), the energy 
flux $P$ and the energy transfer $K$ become negative for high $\omega$. 
}
\end{figure}
Again as in Fig.3(a), the particle flux in Fig.6(a) is negative at small 
$\omega$ and positive above the driving range, 
and it reaches zero both for high and for low $\omega$. 
Compared to the weakly driven system of Fig.3, the particle fluxes 
are stronger by three orders of magnitude, and a higher share of particles 
(one fourth compared to one twentieth) moves toward high $\omega$. 
The relative strength of the direct cascade 
leads to a significant change of the energy fluxes in Fig.6(b): 
The dissipation at high $\omega$ causes to a decay of the energy flux 
$P$ below zero. The dissipation of coupling energy exceeds 
its input by $\Delta E_2$ 
within each cycle of driving and damping. 
Contrary to this 
apparent loss of coupling energy, 
$E_2$ is constant with only small fluctuations, 
and consequently there must be an additional source of coupling energy to 
compensate this loss. \\
This source also plays a role for the energy transfer $K$. 
$K$ is positive in the lower part of the window of 
transparency, which indicates an energy transfer from the inverse cascade 
that is feeding the energy drift in the direct cascade. However, 
the sign of $K$ changes within the window of transparency, and $K<0$ 
suggests that energy is transfered from a source at high $\omega$ 
to lower frequencies, where it feeds the energy drift $Q\omega$. 
The energy drift is 
therefore now driven both by the energy transfer from the inverse cascade, 
and from a transfer of energy from some other source in the high-$\omega$ 
dissipation range. We will now study this source of energy. 
\subsection{Nonlinear energy}
Fig.7 shows the average input and output of particles and of 
nonlinear energy $E_4$ as a function of the amplitude $|\phi|$. 
For a formal definition, we define the ceiling function 
\[
c(v,w)=
\Bigg\{
\begin{array}{lll}
v &\mbox{ for }& v<w \\
w &\mbox{ for }& v\ge w
\end{array}
\]
for real $v$, $w$. 
The square norm of the ceiling function of the amplitude 
$|\tilde{\phi}(x)|$ is an amplitude-dependent particle number 
$n(|\phi|)=\int c^2(|\tilde{\phi}(x)|,|\phi|) dx$. 
This particle number is obtained when the amplitude is set to equal $|\phi|$ 
at all locations where it exceeds $|\phi|$. 
The derivative $n_{|\phi|}=\partial n(|\phi|)/ \partial |\phi|$
gives the nonnegative 
density of particles as a function of the amplitude $|\phi|$, 
and $n=\int_0^{\infty}n_{|\phi|}d|\phi|$ is the total particle number. 
$\Delta n_{|\phi|}$ as shown in Fig.7(a),(b) is the time-averaged difference 
of $n_{|\phi|}$ before and after the driving and damping kick. 
In other words, $n_{|\phi|}$ is the particle density in amplitude space 
that is injected or removed by the driving 
and damping. 
\begin{figure}[b]
\epsfbox{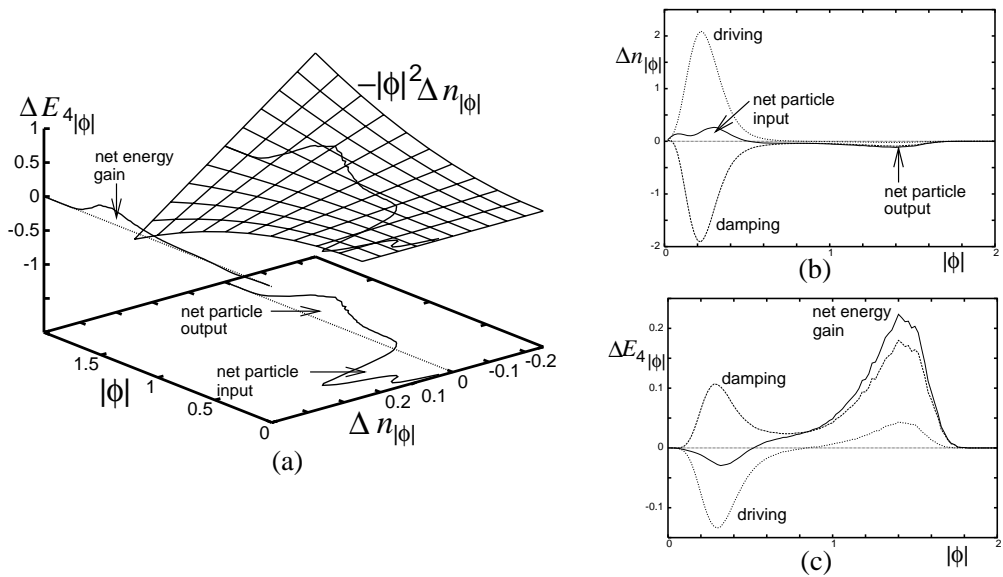}
\caption{Balance of particles $n$ and nonlinear energy $E_4$ as a function 
of the amplitude for the strongly driven system of Fig.2. (a): 
Damping and driving lead to a net particle gain at those locations 
in real space where the amplitude is small ($|\phi|\approx 0.3)$ 
and to a net particle loss at higher amplitudes  
($|\phi|\approx 1.4)$. The net losses match the net gains. 
The losses at high $|\phi|$ lead to a gain of nonlinear energy 
$-|\phi|^2\Delta n_{|\phi|}$. 
(b): Driving and damping inject and remove particles mainly 
at low $|\phi|$. The net particle change (see also (a)) 
is the sum of the driving 
and the damping process. 
(c): The net change of the energy $E_4$ 
(see also (a)) as the sum of the driving 
and the damping process.
}
\end{figure}
There is a net particle input at low amplitudes and a 
net particle output at high amplitudes. The area under the curve where 
particles are gained equals the area under the curve where particles 
are removed, as there is no net gain or loss of particles. 
\\
Analogously to the particle density, a density of nonlinear
energy in amplitude space 
${E_4}_{|\phi|}=\partial E_4(|\phi|)/\partial(|\phi|)$
derives from the amplitude-dependent energy  
$E_4(|\phi|)=-\int c^4(|\tilde{\phi}(x)|,|\phi|)/2 dx$.  
$\Delta {E_4}_{|\phi|}$ of Fig.7(a),(c) is the average difference 
of ${E_4}_{|\phi|}$ before and after the driving and damping. 
This shows a net gain of nonlinear energy at high $|\phi|$ where 
particles are dissipated, and a small net energy loss at low $|\phi|$ 
where particles are injected. 
As the change of the amplitude 
$\Delta |\phi|\ll |\phi|$ is small compared to the amplitude, 
the change of the energy density is connected to the change of the 
particle density by $\Delta {E_4}_{|\phi|}\approx -|\phi|^2\Delta n_{|\phi|}$. 
Sources and sinks of the 
particles translate to sinks and sources of nonlinear energy 
via the factor $-|\phi|^2$. 
This is somewhat similar to the correspondence between sources 
of particles and coupling energy in 
$k$-space of Fig.3, where the coefficient is $\omega(k)$. 
The particle sink at high $|\phi|$ 
becomes a substantial source of energy $E_4$, as the factor $-|\phi|^2$ 
is quite significant there. At the same time, the particle source at low 
$|\phi|$ is only a small energy sink.\\
Figs.7 (b) and (c) distinguish between the effects of driving, damping, 
and the net effect of the superposition of both (as in (a)). 
Fig.7 (b) shows that that most particles are both injected 
and dissipated at locations where the amplitude is small. 
The sum of both curves shows again 
the net particle input at $|\phi|\approx 0.3$ and the output at 
$|\phi|\approx 1.4$. Fig.7(c) shows the corresponding changes  
of nonlinear energy, where the gain at high amplitudes is the most 
important effect. In contrast to this, 
no relevant change of the nonlinear energy is observed in 
the simulation with weak driving of Fig.3. In this case, the particles are 
both injected and removed at low amplitudes, and there is no 
particle dissipation and no energy gain at high $|\phi|$. 
\subsection{Fluxes in amplitude space and energy from collapses} 
The particle source at low $|\phi|$ and the sink at high $|\phi|$ 
induce a particle flux $\Lambda(|\phi|)$ through amplitude space. 
Fig.8(a) shows this particle flux, which is obtained from 
$\partial \Lambda(|\phi|)/\partial |\phi|-\Delta n_{|\phi|}/\Delta t=0$ 
by integration of $\Delta n_{|\phi|}$ over $|\phi|$. 
Similarly to the particle fluxes $Q$ in 
$\omega$-space (Figs.3(a),4(a)), it is zero both for low and for 
high $|\phi|$, as the total particle number is constant. 
$\Lambda(|\phi|)$ is positive, which shows that the particle 
flux is directed from low amplitudes to high amplitudes. 
\begin{figure}[t]
\epsfbox{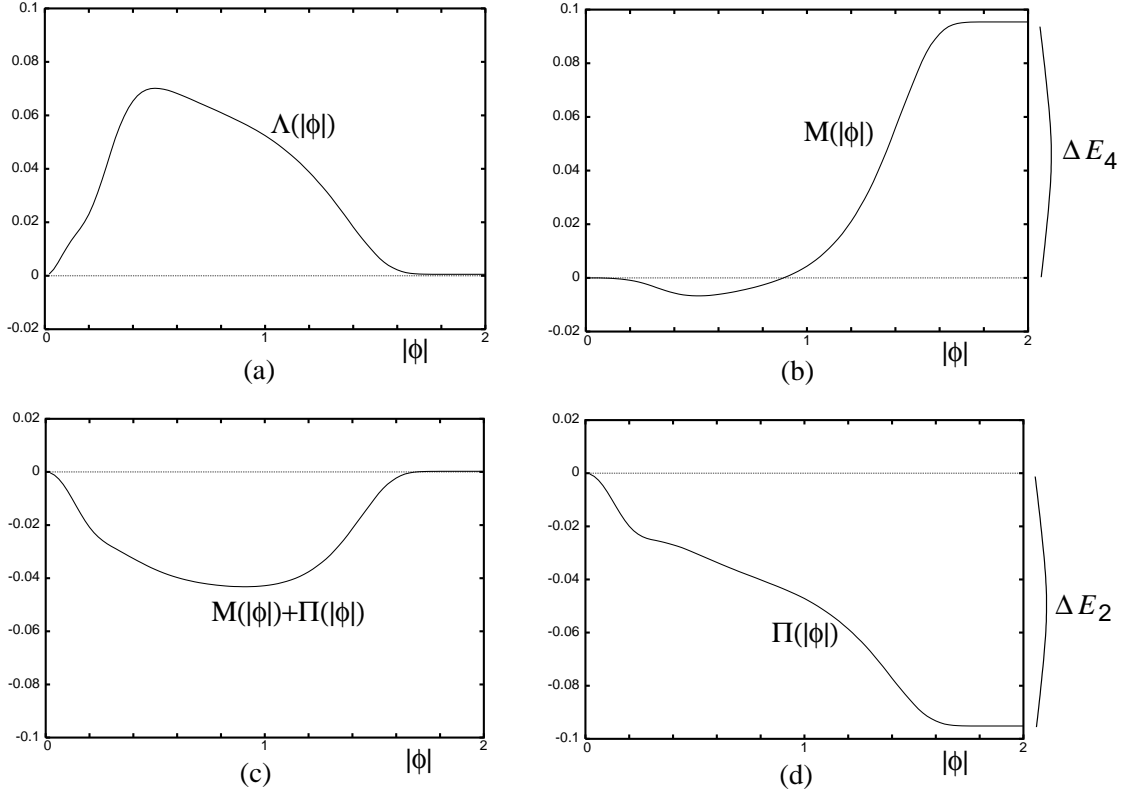}
\caption{Particle and energy fluxes in 
amplitude ($|\phi|$) space. 
(a) Particle flux $\Lambda(|\phi|)$ as obtained by integration 
of the net particle balance of Fig.7. Particles move towards 
high $|\phi|$ as coherent structures grow. $\Lambda (|\phi|)$ 
is zero both for $|\phi|=0$ and for high $|\phi|$, as the system 
neither gains nor loses particles. (b) The "flux" 
$\mbox{M}(|\phi|)$ of nonlinear energy. 
(c) The total flux of energy (nonlinear energy $\mbox{M}(|\phi|)$ plus 
coupling energy $\Pi(|\phi|)$) is negative, i.e. energy flows from 
high amplitudes to low amplitudes. 
(d) The "flux" of coupling energy 
$\Pi(|\phi|)$ is negative. The loss of coupling energy $\Delta E_2$ 
in the flux $\Pi(|\phi|)$ is identical with the loss of coupling 
energy in the flux $P$ in Fig.4. 
The loss of coupling energy $\Delta E_2$ matches the gain 
$\Delta E_4$ of nonlinear energy. 
}
\end{figure}
Unlike the fluxes in $k$ or $\omega$-space, the sources and sinks overlap, 
so that only a part of the total particle input contributes to the 
flux, while most particles are again removed by the damping at the 
same $|\phi|$. \\
A nonlinear energy "flux" 
$\mbox{M}(|\phi|)$ is obtained by integrating 
$\partial \mbox{M}(|\phi|)/\partial |\phi|-\Delta {E_4}_{|\phi|}/\Delta t=0$. 
This quantity is negative at small amplitudes, since the particle sources at low $|\phi|$ 
are sinks of $E_4$. At higher $|\phi|$ it increases with the amplitude, 
since the particle sinks in this domain are powerful sources of $E_4$. 
It reaches a saturation value 
$\Delta E_4$ at $|\phi|\approx 1.8$, which is the  
net gain of nonlinear energy by the driving and the 
damping. 
\\
In a similar way, a "flux" $\Pi(|\phi|)$ of coupling energy 
in amplitude space can be defined. 
The coupling energy of the locations $x$ where the amplitude 
$\tilde{\phi}(x)$ is below the threshold 
$|\tilde{\phi}(x)|\le|\phi|$ is 
$E_2(|\phi|)=\int u(|\phi|-|\tilde{\phi}(x)|)|
|\partial/\partial x|^{1/4}\tilde{\phi}(x)|^2 dx$, 
where $u(l\ge 0)=1, u(l<0)=0$ is the unit step function. 
The corresponding energy density in amplitude space is 
${E_2}_{|\phi|}=\partial E_2(|\phi|)/\partial |\phi|$. 
We measure the time-averaged energy input or output 
$\Delta {E_2}_{|\phi|}$ of the density 
of the coupling energy in amplitude space. 
$\Pi(|\phi|)$ is obtained by integrating 
$\Delta {E_2}_{|\phi|}$ over $|\phi|$ (Fig.8(d)). 
$\Pi(|\phi|)$ is zero at $|\phi|=0$, 
and decreases to a constant value $\Delta E_2<0$ at $|\phi|\approx 1.8$. 
$\Pi(|\phi|)$ expresses in amplitude space the same loss of coupling 
energy as Fig.6(b) does for the flux $P(\omega)$ 
in frequency space. This loss of coupling energy $\Delta E_2$ matches 
the gain of nonlinear energy $\Delta E_4$ of Fig.8(b). \\
These gains and losses 
are caused by the external driving and damping, but the Hamiltonian 
dynamics reverses these changes. 
$E_4$ and $E_2$ each are not conserved 
under the Hamiltonian dynamics, which actually transforms nonlinear 
energy into coupling energy at the same rate as damping 
and driving feed in nonlinear energy, and dissipate coupling energy. 
As a consequence, $E_4$ and $E_2$ each have no gains or losses  
after an average cycle of driving, damping, and Hamiltonian evolution 
over one time unit. 
Nonlinear energy is produced mainly by the dissipation of particles 
at locations where the amplitude is high. The dissipation of coupling energy 
is also relevant for low-amplitude fluctuations. 
The total energy flux $\mbox{M}(|\phi|)+\Pi(|\phi|)$ (Fig.8(c)) 
is therefore directed from high amplitudes to low amplitudes. 
In wavenumber space, this is related to the negative energy transfer 
$K(\omega)$ at high $\omega$ (Fig.6(b)). $K$ is fed by 
the source of nonlinear energy, which is due to the dissipation at high 
frequencies. This energy is 
transformed into coupling energy in 
the window of transparency. \\
$\mbox{M}(|\phi|)$, 
$\Pi(|\phi|)$ and $P(\omega)$ are not genuine fluxes, because 
$E_2$ and $E_4$ each are not conserved under 
the Hamiltonian dynamics. This causes the false impression 
of Figs.6(b),8(b),8(d) of permanent losses of $E_2$ and gains 
of $E_4$. These quantities do not account for the $k$-nonlocal 
energy transformation of $E_4$ into $E_2$. 
In contrast, $\Lambda(|\phi|)$ and $Q(\omega)$ as well as 
$\mbox{M}(|\phi|)+\Pi(|\phi|)$ are fluxes of the conserved 
quantities $n$ and $E$ respectively. In the weakly driven system (Fig.4(b)), 
$P(\omega)$ may also be considered as a flux, because $E_2$ is 
virtually constant under the Hamiltonian dynamics. 
\subsection{The energy cycle of collapses}
We now investigate the Hamiltonian process that transforms energy 
$E_4$ into $E_2$ by studying the time-evolution, 
instead of averaged quantities. 
Figs.9,10 show a simulation with a 
medium-power driving force with $c=13.6$. 
Fig.9(a) shows the maximum square amplitude $max(|\phi|^2(x))$ that 
is reached at any location $x$ in the system. The system has almost 
everywhere low amplitudes $|\phi|^2<0.3$, and high amplitude structures 
$|\phi|^2>1$ emerge only intermittently 
in small areas in space. 
The medium-power driving force allows us to distinguish different 
amplitude blow-ups in time, while a stronger driving force would lead 
to overlapping high-amplitude signals from simultaneous blow-ups 
at different locations. The fluxes of the 
the cases $c=13.6$ and $c=27$ (Figs.6-8) have similar qualitative features. 
The weak driving force with $c=2.4$ of Fig.4 leads to no such 
high-amplitude events, 
so that the maximum amplitude is of the order $|\phi|^2\sim 0.01$. 
\\
It is not completely clear how these structures emerge from the random 
low-amplitude background. 
The nonlinear events are very similar to 
collapses in Hamiltonian partial differential equations 
such as the undamped MMT equation or the nonlinear Schr\"odinger equation. 
These can be described as heteroclinic orbits that connect the 
phase-unstable homogeneous mode with a localized structure with an infinite 
amplitude. A collapse contains a finite amount of particles with 
a zero energy, since its nonlinear energy is minus the coupling energy. 
Collapses may be regarded as a coherent and $k$-nonlocal mechanism 
that transports particles towards 
high wavenumbers \cite{dnpz}, \cite{zgpd}, \cite{zdp}. 
Dissipation of short waves prevents the singularity, and leads to 
a decay of the peak. 
\\
In the focusing nonlinear 
Schr\"odinger equation, all plane waves are phase-unstable 
under slow modulations, 
but this mechanism is different in the MMT equation. 
A perturbation $\delta\phi=\exp(ikx)(a(t)\exp(iqx)+b(t)\exp(-iqx))$ 
of a plane wave $\phi=\sqrt{n} \exp(i(kx-(\omega -n)t)$ 
has an eigenvalue with a positive 
real part if $(2n-\nu)\nu>0$ with
$\nu=(\omega(k+q)+\omega(k-q)-2\omega(k))/2$, which requires 
$0<\nu<2n$. For the focusing nonlinear Schr\"odinger equation, this is always 
fulfilled for small $q$ since $\omega=k^2$ is convex and $\nu=q^2>0$. 
\begin{figure}[t]
\epsfbox{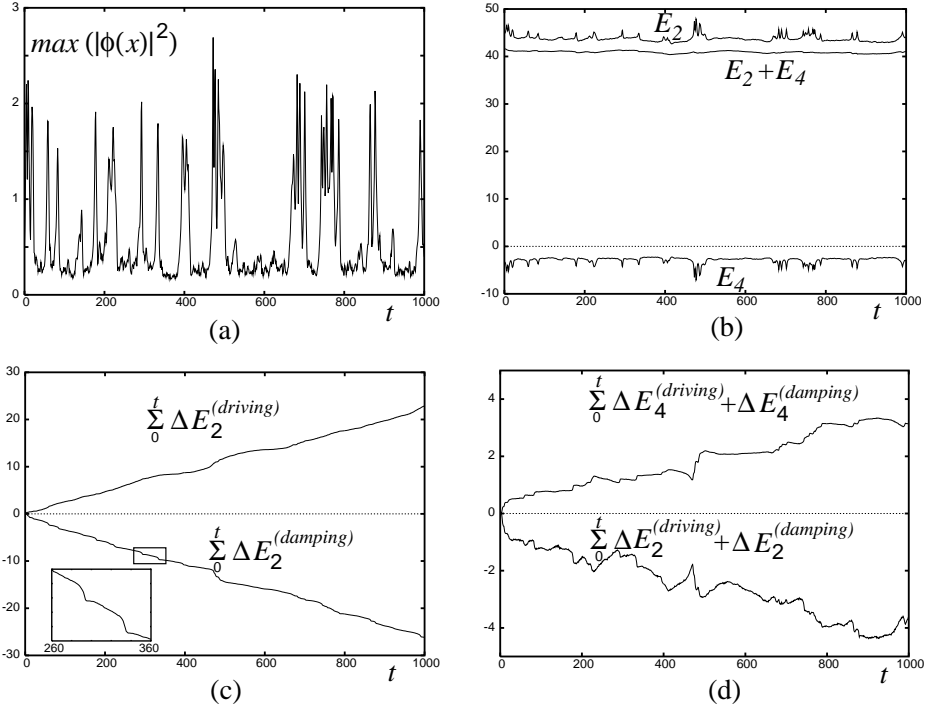}
\caption{Simulation with a medium-power driving force ($c=13.6$). 
(a): Maximum of the squared amplitude $|\phi|^2$ that is 
reached anywhere in the system as a function of time. (b) $E_2$, $E_4$ 
and $E_2+E_4$ as functions of time. (c) Cumulated input of coupling 
energy by the driving force $\sum_0^t \Delta E_2^{(driving)}$, 
and cumulated output of coupling energy by the damping 
$\sum_0^t \Delta E_2^{(damping)}$. (d) Cumulated change of coupling 
energy by the damping and the driving force 
$\sum_0^t \Delta E_2^{(driving)}+\Delta E_2^{(damping)}$ 
(sum of the curves of (c)), and 
cumulated change of nonlinear 
energy by the damping and the driving force 
$\sum_0^t \Delta E_4^{(driving)}+\Delta E_4^{(damping)}$. 
}
\end{figure}
For the MMT equation, 
a necessary condition for an instability is that $k+q$ and $k-q$ 
have different signs, 
i.e. $|q|>|k|$. The homogeneous mode $k=0$ is unstable under slow modulations, 
since the dispersion $\omega=\sqrt{|k|}$ is convex at this point. For 
waves with $k\neq 0$, the dispersion is concave, and an instability 
occurs only by interaction with remote modes, i.e. modes on both the left and 
the right branch of $\omega$. \\
A possible starting point for collapses  might be the built-up 
of a condensate at $k=0$ by the inverse cascade \cite{dnpz}. 
The phase instability of the condensate might lead to a collapse as soon as 
enough particles have gathered at $k=0$. 
However, in our simulations the low-$k$ damping is so efficient that  
the homogeneous mode has a very low level. 
Collapses in our simulations occur while the homogeneous mode contains 
only $n_{k=0}\sim 0.1$ particles, i.e. the number of particles in the 
homogeneous mode of the whole system is below the particle number of a single 
collapse. In addition, we find no sign of a build-up of a condensate, as 
$n_{k=0}$ appears to be independent of the occurrence of collapses. 
Collapses occur in the same way for even more 
efficient low-$k$ damping, 
where each damping step removes all particles with small wavenumbers. 
The collapse dynamics appears to depend crucially on the maximum power in the 
driving range, or equivalently, on the total particle number. 
Instabilities of waves in the driving range are strongly 
nonlocal in wavenumber space and cause no slow phase modulations. 
\\
Fig.9(b) shows the energies $E_2$, $E_4$, and $E=E_2+E_4$ of the total 
system during the simulation. At each of the high-amplitude events, 
$E_4$ temporarily decreases, while $E_2$ increases by the same amount, 
and both energies approach their previous values subsequently. There is 
no net change of these energies on average. \\
Fig.9(c) shows the 
input and output of coupling energy as functions of time. 
$\sum_0^t\Delta E_2^{(driving)}$ is the input of coupling energy through 
the driving force cumulated over a period of time $[0,t]$, 
and $\sum_0^t\Delta E_2^{(damping)}$ is the corresponding output 
through the damping. Coupling energy is dissipated at an approximately 
constant rate between the collapses, which reflects a weakly turbulent 
transport of energy into the dissipation range between 
the collapses. The increased dissipation 
at times when high-amplitude structures emerge leads to 
step-like decays of $\sum_0^t\Delta E_2^{(damping)}$. 
\begin{figure}[t]
\epsfbox{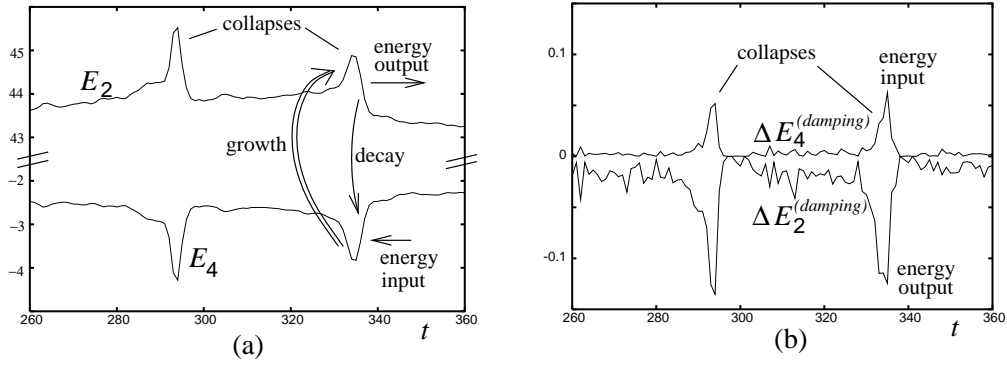}
\caption{
(a): $E_2$ and $E_4$ as functions of time for the simulation of Fig.9. 
Energy is transfered from $E_4$ to 
$E_2$ during the growth of a peak. This energy 
is partly dissipated and partly transfered back 
to $E_4$ when the peak decays. The damping increases the energy 
$E_4$. (b): Energy gain $\Delta E_4^{(damping)}$ and energy loss 
$\Delta E_2^{(damping)}$ by the damping. 
}
\end{figure}
The steps lead to an energy output that is 
bigger than the energy input from the driving, i.e. 
$\sum_0^t\Delta E_2^{(damping)}+\sum_0^t\Delta E_2^{(driving)}<0$. 
This net output of coupling energy is shown in Fig.9(d) together with 
the net input of nonlinear energy 
$\sum_0^t\Delta E_4^{(damping)}+\sum_0^t\Delta E_4^{(driving)}>0$. 
The time average of $\Delta E_2^{(damping)}+\Delta E_2^{(driving)}$ is 
just the energy-flux $P(\omega)=\Pi(|\phi|)=\Delta E_2$ at high $\omega$ or 
high $|\phi|$. 
The average of $\Delta E_4^{(damping)}+\Delta E_4^{(driving)}$ 
is the energy-flux $\mbox{M}(|\phi|)=\Delta E_4$ at high $|\phi|$. 
As there is no net change of the total energy, 
the gain of $E_4$ matches the loss of $E_2$.\\
Fig.10 shows two of the collapses of the simulation of Fig.9. 
When a peak emerges, the total coupling energy $E_2$ increases 
due to the high gradient of the peak. Simultaneously, the 
nonlinear energy $E_4$ decreases because of the locally high amplitude. 
The blow-up transfers energy from $E_4$ to $E_2$. 
\\
The peak is strongly affected by the damping 
so that it looses coupling energy. Fig.10(b) shows this dissipation, where 
the spikes of the dissipation rate $\Delta E_2^{(damping)}$ correspond 
to the steps of $\sum_0^t\Delta E_2^{(damping)}$ of Fig.9(c). 
As the amplitude of the peak decreases 
by the action of the damping, the nonlinear energy increases again. 
Therefore the damping provides both an input of energy $E_4$ and an output 
of energy $E_2$ (Fig.10(b)). The picture of the energy cycle is that 
the damping feeds 
nonlinear energy $\Delta E_4$ into the system, that is transformed to coupling 
energy $\Delta E_2$ and subsequently dissipated. 
The burnouts of the peaks are incomplete \cite{dnpz}, and some of the 
coupling energy is not dissipated, but returned to the nonlinear 
energy by a Hamiltonian process. Radiation of the decaying peak might 
also contribute to the low-amplitude waves. 
\\
Fig.11(a) shows the net output rate coupling energy 
$\Delta E_2$ and the net input rate of nonlinear energy $\Delta E_4$ 
that is caused by damping and driving.  As in Fig.5, $c$ measures the 
particle density that is obtained in the window of transparency for 
different driving powers. 
Again, we can distinguish between 
pure weak turbulence and collapses. 
\begin{figure}[t]
\epsfbox{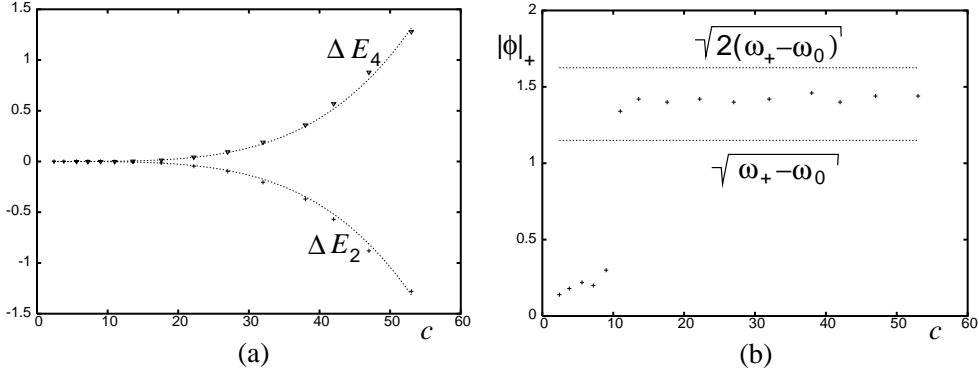}
\caption{(a) Net loss of coupling energy $\Delta E_2$ and 
gain of nonlinear energy 
$\Delta E_4$ versus the coefficient c 
of the power law $n_k=c/(kL/2\pi)$. $\Delta E_2$ is the value of 
the energy flux $P$ at high $\omega$ or $\Pi$ at high $|\phi|$, 
$\Delta E_4$ is the value of the flux 
of nonlinear energy $\mbox{M}$ at high $|\phi|$. 
(b) Peak heights together with an upper and a lower estimate. 
The simulations are the same as in Fig.5. 
}
\end{figure}
For driving forces below the threshold $c\approx 10$, the cumulated  
dissipation of coupling energy (as in Fig.9(c)) has no steps, as there are no 
collapses. Average input and output of coupling energy are matching as 
$\Delta E_2^{(damping)}+\Delta E_2^{(driving)}=\Delta E_2\approx 0$, and 
the input of nonlinear energy 
$\Delta E_4^{(damping)}\approx 0$ and 
$\Delta E_4^{(driving)}\approx 0$ is negligible. Above 
the threshold, output of coupling energy and input of nonlinear energy 
by damping and driving 
match closely as $\Delta E_2=-\Delta E_4$. 
The lines give a tentative fit $\sim c^4$. 
\\
A characteristic peak height can be defined as the amplitude 
where the nonlinear energy change by the damping and driving  is maximal, 
i.e. $|\phi|$ of Fig.7 with the maximum $|\Delta {E_4}_{|\phi|}|$. 
Fig.11(b) shows this peak height $|\phi|_+$ versus the coefficient $c$ 
of the power spectrum. Again, high amplitudes appear only for 
strong driving forces. Below the threshold, $|\Delta {E_4}_{|\phi|}|$ 
is very small and negligible in the energy balance. 
To give an estimate the height of these structures, we 
assume that $\Delta n$ particles 
carrying an energy $\omega_{0}\Delta n$ are injected at $\omega_{0}$, 
and they are removed at $\omega_{+}$ where they carry the coupling energy 
$\omega_{+} \Delta n$. The net energy loss 
$\Delta E_2=(\omega_{0}-\omega_{+}) \Delta n$ 
from this process can be balanced by a gain of nonlinear energy due 
to a particle flow towards high amplitudes where the particles are
dissipated. If we assume for simplicity that the damping 
reduces the amplitude $|\phi|$ by $\Delta|\phi|\ll |\phi|$ 
in the interval $\Delta x$ in space,  an amount of 
$\Delta n=2|\phi|\Delta |\phi|\Delta x$ particles are removed and 
the nonlinear energy is increased by $\Delta E_4=|\phi|^2 \Delta n$. 
With the energy balance  $\Delta E_4+\Delta E_2=0$, we obtain  
$|\phi|=\sqrt{\omega_{+}-\omega_{0}}$ as a lower estimate for the 
height of the peaks. 
For an upper estimate, we assume that the damping decreases the amplitude by 
$\Delta|\phi|= |\phi|$ and removes all particles 
$\Delta n=|\phi|^2 \Delta x$ from this interval, so that the 
gain of nonlinear energy is 
$\Delta E_4=|\phi|^4\Delta x/2=|\phi|^2\Delta n/2$. 
For this case, the energy balance gives 
$|\phi|=\sqrt{2(\omega_{+}-\omega_{0})}$ as 
an upper estimate of the peak height. 
If the particles input by the driving force has an average 
$\omega_{0}\approx 0.2$ and the particle output occurs 
at $\omega_+\approx 1.5$, we expect a peak height between $|\phi|\approx1.1$
and $|\phi|\approx 1.6$, compared to the numerical result $|\phi|\approx
1.4$. 
\section{Conclusions}
We have studied the fluxes of energy and particles in a onedimensional 
turbulent system, the focusing Majda-McLaughlin-Tabak equation. These fluxes 
reveal details about the statistically stationary state 
that cannot be observed directly in the conserved quantities, 
such as the spectrum of particles $n_k$ (Fig.2). 
Fluxes in frequency space (Figs.4,6) have sources 
and sinks that are determined by the external driving and damping. 
Fluxes in amplitude space (Fig.8) emerge spontaneously 
when the total particle input rate exceeds a threshold (Fig.5). 
\\
We have found two types of  nonequilibria with 
similar spectra (Fig.2(a)), but very different energy fluxes. 
The nonequilibrium state for sufficiently weak driving 
is very similar to the weakly turbulent Zakharov solution. 
The average amplitude is small, 
with an exponentially decreasing probability for higher amplitudes (Fig.2(b)). 
Nonlinearity only plays a role as a weak coupling between the dispersive 
waves, and no high-amplitude structures emerge. 
Particles that are injected at the driving scale flow toward the 
sinks both at high and at low wavenumbers (Fig.4(a)). 
We interpret the well-known quantities \cite{zlf} $Q\omega$ 
and $K$ (Fig.4(b)) as energy drift and energy transfer. 
In particular, $K$ transfers energy that is released in 
the inverse cascade to the direct cascade, where it is is absorbed. 
The fluxes increase slightly faster with the particle density (Fig.5(a)) 
than predicted by the theory of weak turbulence. \\
For stronger driving forces, high amplitude structures emerge intermittently, 
and the interaction becomes strongly nonlinear. 
The energy balance changes in two ways: Firstly, more coupling energy is 
dissipated by the damping than supplied by the driving force. 
This leads to a negative energy flux $P$ and energy transfer $K$ 
for high frequencies (Fig.6(b)). Secondly, 
damping and driving provide a net surplus 
of nonlinear energy $E_4$, which is due to 
the dissipation of particles at high amplitudes (Figs.7,8(b)). 
This is related to fluxes of particles 
and energy in amplitude space (Fig.8). There is a net particle flux from 
low amplitudes, where the driving force injects particles, toward high 
amplitudes, where particles are removed by the dissipation (Fig.8(a)). 
The dissipation of particles at the tip of high-amplitude structures 
is the source of an energy flux from high amplitudes to low amplitudes, 
where coupling energy is dissipated (Fig.8(c)). \\
While driving and damping cause a permanent 
net input of $E_4$ and an equal net output 
of $E_2$, both the coupling energy and the nonlinear energy 
are constant with only small fluctuations (Fig.9(b). 
This is possible because the 
Hamiltonian dynamics transforms nonlinear energy into coupling 
energy within the energy flux from high to low amplitudes (Fig.8(c)). 
This transformation of nonlinear energy into coupling energy occurs 
during the blow-ups (Figs.9,10), where the coupling energy increases 
and the nonlinear energy decreases. Coupling energy can 
be radiated into the surrounding low-amplitude waves, 
when, for example, a peak emerges 
out of two colliding solitary waves \cite{dzpsy}. 
A peak that decays after reaching its maximum height can also radiate 
low-amplitude waves. The generation of coupling energy at high $\omega$ 
feeds the energy transfer $K$, that is directed toward low frequencies 
at high $\omega$. The representation of energy 
fluxes in frequency or wavenumber space is 
somehow misleading since it neglects the $k$-nonlocal 
transformation nonlinear energy into coupling energy.  
\\
The nonlinear energy $E_4$ is related to high-amplitude coherent 
structures, while the coupling energy $E_2$ has contributions 
both from these peaks and from low-amplitude fluctuations. The 
advantage of studying these two quantities (and not peaks and fluctuations) 
is that their fluxes are measurable statistical quantities that 
distinguish weakly turbulent and coherent processes. 
The fluxes in wavenumber or frequency space and in amplitude space 
may be seen as two projections of the complex highdimensional 
transport process. In wavenumber space, driving and damping 
is related to localized sources and sinks of the fluxes. 
The fluxes yield the boundary conditions for the 
nonequilibrium in the inertial range. 
This is appropriate for the pure weakly turbulent process, 
where the statistics is almost Gaussian and higher order correlations 
can be reduced to two-point functions. If the statistics is non-Gaussian, 
the external forces inject energy to the nonlinear part of 
the Hamiltonian $E_4$ (i.e. the four-point correlation) 
and remove energy from $E_2$, the two-point correlation. The Hamiltonian 
dynamics is characterized by the transformation of energy $E_4$ into $E_2$, 
which is actually the transfer of a conserved quantity between 
correlations of different order.  
The representation in amplitude 
space covers this spontaneous coherent process, 
which appears as an energy flux from high amplitudes to low amplitudes. 
The representations in wavenumber space and in amplitude space together 
cover the energetics of this process. 
The many open questions 
on the energy transfer from $E_4$ to $E_2$ 
concern, in particular, the onset of this coherent behavior. 
Also, the effect of the  
coherent structures on the low-amplitude process by radiation is not 
well understood.

\end{document}